\documentclass[10pt, twocolumn, final]{IEEEtran}%
\usepackage[T1]{fontenc}
\usepackage[utf8]{inputenc}
\usepackage{amsmath,amssymb,amsfonts,mathrsfs,bm}% Typical maths resource packages
\usepackage{mathtools}
\usepackage{amsthm}
\usepackage{nicefrac}
\usepackage[shortlabels]{enumitem}
\usepackage{graphicx}
\usepackage{epstopdf}
\usepackage{url}
\usepackage{colortbl}
\usepackage{booktabs}
\usepackage{multirow}
\usepackage[table,dvipsnames]{xcolor}
\usepackage[normalem]{ulem}
\usepackage{xparse}

\makeatletter
\@ifpackageloaded{natbib}{
	\relax
}{
	\usepackage{cite}
}
\makeatother

\usepackage{array}
\newcolumntype{L}[1]{>{\raggedright\let\newline\\\arraybackslash\hspace{0pt}}m{#1}}
\newcolumntype{C}[1]{>{\centering\let\newline\\\arraybackslash\hspace{0pt}}m{#1}}
\newcolumntype{R}[1]{>{\raggedleft\let\newline\\\arraybackslash\hspace{0pt}}m{#1}}

\makeatletter
\let\MYcaption\@makecaption
\makeatother
\usepackage[font=footnotesize]{subcaption}
\makeatletter
\let\@makecaption\MYcaption
\makeatother



\usepackage{glossaries}

\makeatletter
\let\oldgls\gls
\let\oldglspl\glspl

\newcommand\fussy@ifnextchar[3]{%
	\let\reserved@d=#1%
	\def\reserved@a{#2}%
	\def\reserved@b{#3}%
	\futurelet\@let@token\fussy@ifnch}
\def\fussy@ifnch{%
	\ifx\@let@token\reserved@d
		\let\reserved@c\reserved@a
	\else
		\let\reserved@c\reserved@b
	\fi
	\reserved@c}

\renewcommand{\gls}[1]{%
\oldgls{#1}\fussy@ifnextchar.{\@checkperiod}{\@}}
\renewcommand{\glspl}[1]{%
\oldglspl{#1}\fussy@ifnextchar.{\@checkperiod}{\@}}

\newcommand{\@checkperiod}[1]{%
	\ifnum\sfcode`\.=\spacefactor\else#1\fi
}
\makeatother

\newacronym{wrt}{w.r.t.}{with respect to}
\newacronym{RHS}{R.H.S.}{right-hand side}
\newacronym{LHS}{L.H.S.}{left-hand side}
\newacronym{iid}{i.i.d.}{independent and identically distributed}
\usepackage{float}

\ifx\notloadhyperref\undefined
	\ifx\loadbibentry\undefined
		\usepackage[hidelinks,hypertexnames=false]{hyperref}
	\else
		\usepackage{bibentry}
		\makeatletter\let\saved@bibitem\@bibitem\makeatother
		\usepackage[hidelinks,hypertexnames=false]{hyperref}
		\makeatletter\let\@bibitem\saved@bibitem\makeatother
	\fi
\else
	\ifx\loadbibentry\undefined
		\relax
	\else
		\usepackage{bibentry}
	\fi
\fi

\usepackage[capitalize]{cleveref}
\crefname{equation}{}{}
\Crefname{equation}{}{}
\crefname{claim}{claim}{claims}
\crefname{step}{step}{steps}
\crefname{line}{line}{lines}
\crefname{condition}{condition}{conditions}
\crefname{dmath}{}{}
\crefname{dseries}{}{}
\crefname{dgroup}{}{}

\crefname{Problem}{Problem}{Problems}
\crefformat{Problem}{Problem~(#2#1#3)}
\crefrangeformat{Problem}{Problems~(#3#1#4) to~(#5#2#6)}

\crefname{Theorem}{Theorem}{Theorems}
\crefname{Corollary}{Corollary}{Corollaries}
\crefname{Proposition}{Proposition}{Propositions}
\crefname{Lemma}{Lemma}{Lemmas}
\crefname{Definition}{Definition}{Definitions}
\crefname{Example}{Example}{Examples}
\crefname{Assumption}{Assumption}{Assumptions}
\crefname{Remark}{Remark}{Remarks}
\crefname{Rem}{Remark}{Remarks}
\crefname{remarks}{Remarks}{Remarks}
\crefname{Appendix}{Appendix}{Appendices}
\crefname{Supplement}{Supplement}{Supplements}
\crefname{Exercise}{Exercise}{Exercises}
\crefname{Theorem_A}{Theorem}{Theorems}
\crefname{Corollary_A}{Corollary}{Corollaries}
\crefname{Proposition_A}{Proposition}{Propositions}
\crefname{Lemma_A}{Lemma}{Lemmas}
\crefname{Definition_A}{Definition}{Definitions}

\usepackage{crossreftools}
\ifx\notloadhyperref\undefined
\else
	\relax
\fi

\usepackage{algorithm,algorithmic}

\ifx\loadbreqn\undefined
	\relax
\else
	\usepackage{breqn}
\fi

\ifx\renewtheorem\undefined
	\ifx\useTheoremCounter\undefined
		\newtheorem{Theorem}{Theorem}
		\newtheorem{Corollary}{Corollary}
		\newtheorem{Proposition}{Proposition}
		
	\else

	\fi

	\newtheorem{Definition}{Definition}
	\newtheorem{Example}{Example}

	\newtheorem{Proposition_A}{Proposition}[section]
	\newtheorem{Lemma_A}{Lemma}[section]
	\newtheorem{Definition_A}{Definition}[section]
	\newtheorem{Example_A}{Example}[section]
\fi

\theoremstyle{remark}

\theoremstyle{plain}

\newcommand{\qednew}{\nobreak \ifvmode \relax \else
		\ifdim\lastskip<1.5em \hskip-\lastskip
			\hskip1.5em plus0em minus0.5em \fi \nobreak
		\vrule height0.75em width0.5em depth0.25em\fi}

\NewDocumentCommand{\movedownsub}{e{^_}}{%
	\IfNoValueTF{#1}{%
		\IfNoValueF{#2}{^{}}% neither ^ nor _, do nothing; if no ^ but _, add ^{}
	}{%
		^{#1}% add superscript if present
	}%
	\IfNoValueF{#2}{_{#2}}% add subscript if present
}

\let\latexchi\chi
\RenewDocumentCommand{\chi}{}{\latexchi\movedownsub}

\newcommand{\calI}{\mathcal{I}}

\newcommand{\calO}{\mathcal{O}}

\newcommand{\bA}{\mathbf{A}}

\newcommand{\bD}{\mathbf{D}}

\newcommand{\bF}{\mathbf{F}}

\newcommand{\bH}{\mathbf{H}}

\newcommand{\bI}{\mathbf{I}}

\newcommand{\bL}{\mathbf{L}}

\newcommand{\bM}{\mathbf{M}}

\newcommand{\bP}{\mathbf{P}}

\newcommand{\bS}{\mathbf{S}}

\newcommand{\bU}{\mathbf{U}}

\newcommand{\bW}{\mathbf{W}}

\newcommand{\bX}{\mathbf{X}}

\newcommand{\scJ}{\mathscr{J}}

\DeclareSymbolFont{bsfletters}{OT1}{cmss}{bx}{n}
\DeclareSymbolFont{ssfletters}{OT1}{cmss}{m}{n}
\DeclareMathSymbol{\bsfGamma}{0}{bsfletters}{'000}
\DeclareMathSymbol{\ssfGamma}{0}{ssfletters}{'000}
\DeclareMathSymbol{\bsfDelta}{0}{bsfletters}{'001}
\DeclareMathSymbol{\ssfDelta}{0}{ssfletters}{'001}
\DeclareMathSymbol{\bsfTheta}{0}{bsfletters}{'002}
\DeclareMathSymbol{\ssfTheta}{0}{ssfletters}{'002}
\DeclareMathSymbol{\bsfLambda}{0}{bsfletters}{'003}
\DeclareMathSymbol{\ssfLambda}{0}{ssfletters}{'003}
\DeclareMathSymbol{\bsfXi}{0}{bsfletters}{'004}
\DeclareMathSymbol{\ssfXi}{0}{ssfletters}{'004}
\DeclareMathSymbol{\bsfPi}{0}{bsfletters}{'005}
\DeclareMathSymbol{\ssfPi}{0}{ssfletters}{'005}
\DeclareMathSymbol{\bsfSigma}{0}{bsfletters}{'006}
\DeclareMathSymbol{\ssfSigma}{0}{ssfletters}{'006}
\DeclareMathSymbol{\bsfUpsilon}{0}{bsfletters}{'007}
\DeclareMathSymbol{\ssfUpsilon}{0}{ssfletters}{'007}
\DeclareMathSymbol{\bsfPhi}{0}{bsfletters}{'010}
\DeclareMathSymbol{\ssfPhi}{0}{ssfletters}{'010}
\DeclareMathSymbol{\bsfPsi}{0}{bsfletters}{'011}
\DeclareMathSymbol{\ssfPsi}{0}{ssfletters}{'011}
\DeclareMathSymbol{\bsfOmega}{0}{bsfletters}{'012}
\DeclareMathSymbol{\ssfOmega}{0}{ssfletters}{'012}

\DeclareMathOperator{\st}{s.t.\ }
\DeclarePairedDelimiterX\ip[2]{\langle}{\rangle}{#1,#2}
\DeclarePairedDelimiterX\norm[1]{\lVert}{\rVert}{#1}
\DeclarePairedDelimiterXPP\col[1]{\operatorname{col}}{\{}{\}}{}{#1} % column vector
\DeclarePairedDelimiterXPP\row[1]{\operatorname{row}}{\{}{\}}{}{#1} % row vector
\DeclarePairedDelimiterXPP\erf[1]{\operatorname{erf}}{(}{)}{}{#1}
\DeclarePairedDelimiterXPP\erfc[1]{\operatorname{erfc}}{(}{)}{}{#1}
\DeclarePairedDelimiterXPP\op[2]{\operatorname{#1}}{(}{)}{}{#2} % general operator

\providecommand\given{}
\newcommand\SetSymbol[2][]{%
	\nonscript\, #1#2
	\allowbreak
	\nonscript\,
	\mathopen{}}

\DeclarePairedDelimiterX\Set[2]\{\}{%
\renewcommand\given{\SetSymbol[\delimsize]{#1}}
#2
}
\DeclarePairedDelimiterX\Setc[1]\{\}{%
\renewcommand\given{\SetSymbol{:}}
#1
}

\NewDocumentCommand\set{s o m}{%
	\IfBooleanTF#1%
	{\IfValueTF{#2}{\Set*{#2}{#3}}{\Setc*{#3}}}%
	{\IfValueTF{#2}{\Set{#2}{#3}}{\Setc{#3}}}%
}

\NewDocumentCommand{\evalat}{s O{\big} m m}{%
\IfBooleanTF{#1}
{{\left. #3 \right|_{#4}}}
{{#3#2|_{#4}}}%
}

\NewDocumentCommand \ifcond {m m} {%
	{#1} %
	\IfValueT{#2}{\, \middle|\, {#2}}%
}

\DeclareDocumentCommand \P {e{_} g >{\SplitArgument{ 1 }{ @| }}d() g } {%
	\mathbb{P}%
	\IfValueTF{#1}{_{#1}}
	{\IfValueT{#2}{_{#2}}}%
	\IfValueT{#3}{\left(\ifcond#3}%
	\IfValueT{#4}{\, \middle|\, {#4}}%
	\IfValueT{#3}{\right)}%
}

\DeclareDocumentCommand \E {e{_} g >{\SplitArgument{ 1 }{ @| }}o g } {%
\mathbb{E}%
\IfValueTF{#1}{_{#1}}
{\IfValueT{#2}{_{#2}}}%
\IfValueT{#3}{\left[\ifcond#3}%
\IfValueT{#4}{\, \middle|\, {#4}}%
\IfValueT{#3}{\right]}%
}

\let\oldforall\forall
\renewcommand{\forall}{\oldforall \, }

\let\oldexist\exists
\renewcommand{\exists}{\oldexist \, }

\renewcommand{\figurename}{Fig.}
\newcommand{\figref}[1]{\figurename~\ref{#1}}
\graphicspath{{./Figures/}{./figures/}}
\newcommand{\includeCroppedPdf}[2][]{%
	\IfFileExists{./Figures/#2-crop.pdf}{}{%
		\immediate\write18{pdfcrop ./Figures/#2 ./Figures/#2-crop.pdf}}%
	\includegraphics[#1]{./Figures/#2-crop.pdf}}

\definecolor{gray90}{gray}{0.9}

\ifx\nohighlights\undefined

	\newcommand{\msout}[1]{\text{\color{green} \sout{\ensuremath{#1}}}}
	\newcommand{\del}[1]{{\color{green}\ifmmode \msout{#1}\else\sout{#1}\fi}}
\else

	\newcommand{\msout}[1]{#1}
	\newcommand{\del}[1]{#1}
\fi

\newcommand{\hhide}[1]{}
\ifx\diagnoselabel\undefined
	\relax
\else
	\makeatletter
	\def\@testdef #1#2#3{%
		\def\reserved@a{#3}\expandafter \ifx \csname #1@#2\endcsname
			\reserved@a  \else
			\typeout{^^Jlabel #2 changed:^^J%
				\meaning\reserved@a^^J%
				\expandafter\meaning\csname #1@#2\endcsname^^J}%
			\@tempswatrue \fi}
	\makeatother
\fi

\crefname{question}{question}{questions}

\usepackage{tikz}
\usepackage{color}
\usepackage{pgfplots}
\pgfplotsset{compat=1.5}%%%%%%%%%%%%%%%%%%%

\providecommand{\U}[1]{\protect\rule{.1in}{.1in}}

\theoremstyle{definition}

\newtheorem{condition}{Conditions}

\begin{document}
\date{}
\title{On semi shift invariant graph filters}
\author{Feng~Ji, See Hian Lee, and Wee~Peng~Tay,~\IEEEmembership{Senior Member,~IEEE}%
%\thanks{This research is supported by the Singapore Ministry of Education Academic Research Fund Tier 2 grant MOE2018-T2-2-019 and A*STAR under its RIE2020 Advanced Manufacturing and Engineering (AME) Industry Alignment Fund — Pre Positioning (IAF-PP) (Grant No. A19D6a0053).}% 
%\thanks{The authors are with the School of Electrical and Electronic Engineering, Nanyang Technological University, 639798, Singapore (e-mail: jifeng@ntu.edu.sg, wptay@ntu.edu.sg, giacomo.kahn@gmail.com).}%
}
\maketitle
%\vspace{-10pt}
\begin{abstract}
In graph signal processing, one of the most important subject is the study of filters, i.e., linear transformations that capture relations between graph signals. One of the most important families of filters is the space of shift invariant filters, defined as transformations commute with a preferred graph shift operator. Shift invariant filters have a wide range of applications in graph signal processing and graph neural networks. A shift invariant filter can be interpreted geometrically as an information aggregation procedure (from local neighborhood), and can be computed easily using matrix multiplication. However, there are still drawbacks to use solely shift invariant filters in applications, such as being restrictively homogeneous. In this paper, we generalize shift invariant filters by introducing and study semi shift invariant filters. We give an application of semi shift invariant filters with a new signal processing framework, the subgraph signal processing. Moreover, we also demonstrate how semi shift invariant filters can be used in graph neural networks.   
\end{abstract}


\section{Introduction} \label{sec:intro}

Since its emergence, the theory and applications of graph signal processing (GSP) have rapidly developed. GSP incorporates geometric properties of a graph in analyzing signals supported on it. The theory covers a wide range of topics with a full array of applications \cite{Shu13, San13, San14, Gad14, Don16, Def16, kipf2017semi, Egi17, Sha17, Gra18, Ort18, Girault2018, JiTay:J19, Suc17, Wan18, JiYanZha:C20}. In this paper, we want to study the aspect of filtering. 

Suppose the size of a graph $G$ is $n$. In the nutshell, a graph signal is a vector in $\mathbb{R}^n$, with each component associated with a node of $G$. A filter is a linear transformation, which can be represented by a matrix in $M_n(\mathbb{R})$. Filtering is ubiquitous in signal processing. For example, it can be used to describe relations between signals in prediction, model how signals change from time to time in time-series analysis, as well as processing node features in graph neural networks (cf.\ \cite{Shu13, Def16, kipf2017semi, Ort18}).

The full space of filters $M_n(\mathbb{R})$ is $n^2$ dimensional. This means that if we want to find an appropriate filter for signal processing tasks, the search space can be oversized. Therefore, there are attempts to construct filter banks, subspaces of $M_n(\mathbb{R})$, with smaller dimensions. One of the most important filter banks in GSP is the space of shift invariant filters. Under favorable condition, a shift invariant filter can be expressed as a polynomial in a chosen graph shift operator, such as the graph adjacency matrix or the Laplacian. Shift invariant operators are widely used as they enjoy a few important properties: 
\begin{enumerate}[a)]
    \item A shift operator is usually interpretable as an aggregation mechanism of information from surrounding nodes of each node in the graph. It approximates well our understanding of many actual physical systems in real applications. 
    \item The dimension of the space of shift invariant filters is at most $n$. It is much smaller than the dimension of $M_n(\mathbb{R})$. This makes shift invariant filters preferred in many learning problems, as we do not want any candidate filter bank containing too many unrealistic filters. 
    \item It is usually computationally efficient to handle shift invariant filters as only matrix multiplications are involved.   
\end{enumerate}
Many subspaces of the space of shift invariant filters, such as small degree shift invariant filters and band-pass filters, play prominent roles in many applications.

However, shift invariant filters have some drawbacks as well. For example, a shift invariant filter is homogeneous in the following sense. For the above mentioned graph shift operators, the way signals are gathered for each node from its neighboring nodes does not change from node to node. Simple calculations shows that the same observation holds for any polynomial, i.e., shift invariant filter, of the shift operator. This feature can be restrictive in applications as we can always encounter non-homogeneous situations. Moreover, the space of shift invariant filters can be insufficient in certain applications considering its relatively small dimension.

In this paper, we introduce the spaces of semi shift invariant filters, which are essentially mixtures of shift invariant filters restricted to different subgraphs of the ambient graph $G$. The concept subsumes the above mentioned filter banks such as shift invariant filter, the full filter space $M_n(\mathbb{R})$ as special cases. Moreover, we shall demonstrate that the newly introduced filter banks can resolve certain drawbacks of the shift invariant filters. 

Our main contributions are as follows:
\begin{itemize}
    \item We define semi shift invariant filters and discuss their properties. We compute dimensions of different families of semi shift invariant filters and discuss their includement relations.
    \item We describe a new graph signal processing framework: subgraph signal processing. The purpose is to perform signal processing without observing signals at all the nodes or even without observing the full graph.
    \item We demonstrate that semi shift invariant filters can also be applied to graph neural networks (GNN). We explain how the concept can be systematically synergized with existing GNN models.
\end{itemize}

The rest of the paper is organized as follows. In \cref{sec:semi}, we give a concise overview of GSP and point out why a generalization of the shift invariant filters might be needed, which leads to the introduction of semi shift invariant filters. In the subsequent subsection, we study properties of the newly introduced filter families. In \cref{sec:subgsp} and \cref{sec:gnn}, we describe applications of semi shift invariant filters with simulation results. In \cref{sec:subgsp}, we describe the subgraph signal processing framework and explain the role played by semi shift invariant filters. In \cref{sec:gnn}, we discuss applications of semi shift invariant filters in GNN, including both homogeneous and heterogeneous semi-supervised node classification problems. We conclude in \cref{sec:con}. 

\emph{Notations:} We use ordinary lower-case letters such as $f, g, x, y$ for graph signals. Capital letters are used for matrices and operators, and they are boldfaced. For a matrix $\bM$, its $i,j$-th entry is denoted by $\bM_{i,j}. $We use $\scJ$ with appropriate subscripts to denote filter spaces. A polynomial is denoted by $Q_d$ with the subscript its degree. The vertex set of a graph is usually denoted by $V$, and examples of its subsets are $V_0, V_2, V_1', V_2'$. We use lower-case letter such as $u,v, v_i$ to denote nodes of a graph. 

\section{Semi shift invariant graph filters} \label{sec:semi}

\subsection{Graph signal processing preliminaries}

In this subsection, we give a brief overview of graph signal processing (GSP). The main purpose is to introduce shift invariant (SI) filters and motivate subsequent subsections. 

Let $G=(V,E)$ be a graph of size $n$. A graph signal $f$ assigns a number $f(v)$ to each node $v\in V$, and the space of graph signals can be identified with $\mathbb{R}^n$. A \emph{graph shift operator (GSO)} $\bS$ is a linear transformation on $\mathbb{R}^n$. In the literature, there are a few preferred candidates of $\bS$ such as the adjacency matrix $\bA_G$ of $G$, the Laplacian $\bL_G$ and their normalized versions $\widetilde{\bA_G}, \widetilde{\bL_G}$. These choices all have the geometric interpretation of aggregating signals from $1$-hop neighborhood for each node. Therefore, graph information is contained in each of them. Moreover, such an $\bS$ is normal, which we assume holds true for any chosen GSO in the sequel. It implies that $\bS$ admits an orthonormal decomposition $\bU\Lambda\bU^*$ with $\Lambda$ a diagonal matrix consisting of eigenvalues of $\bS$. The columns of $\bU$ forms the graph Fourier basis of the \emph{frequency domain} associated with $\bS$. \emph{Graph Fourier transform} (GFT) is nothing but base change w.r.t.\ the Fourier basis, i.e., GFT of $f$ is $\hat{f} = \bU^*f$.  

The operator $\bS$ is also a \emph{graph filter}. In general, a graph filter is a linear transformation in $M_n(\mathbb{R})$. One uses filters to describe relations between different signals. One of the main topics of GSP is the construction of filter banks. Associated with $\bS$ itself, there is the (vector) space of \emph{shift invariant (SI) filters} as those filters $\bF$ commute with $\bS$, i.e., $\bF\circ \bS = \bS \circ \bF$. The notion is analogous to that of shift invariance in classical Fourier theory if $G$ is taken as the direct cyclic graph and $\bS = \bA_G$. If $\bS$ does not have repeated eigenvalues, then each SI filter $\bF$ is a polynomial in $\bS$. By the Cayley-Hamilton theorem, the dimension of space of SI filters is $n$. Consequently, a typical SI filter is a linear combination of monomials $\bS^d$, which aggregates $d$-hop information. However, this loses flexibility as the same linear combination is used across the entire graph. In particular, this is not preferred if we only want to gather $d$-hop information for nodes in a subset $V_0$ of $V$. In the next subsection, we introduce semi shift invariant filters that addresses this issue. 

\subsection{Semi shift invariant graph filters}

In this subsection, we introduce the space of semi shift invariant filters as a generalization of the space of shift invariant filters.

Let $\bS$ be a fixed shift operator for $G$ (of size $n$), whose matrix form we assume to be normal. This is a theoretical requirement. As we have seen, the shift operator $\bS$ dictates how information is being passed on $G$. 

\begin{Definition} \label{defn:lvs}
Consider $V_0 \subset V$ and $d\leq n-1$. Let $\bP_{V_0}: \mathbb{R}^{n} \to \mathbb{R}^{|V_0|}$ be the projection, i.e., for $v\in V_0$, the $v$-component of $\bP_{V_0}f$ is $f(v)$. 

Let $\overline{\bP}_{V_0}: \mathbb{R}^{n}\to \mathbb{R}^{n}$, $f \mapsto \overline{\bP}_{V_0}f$ be such that $\overline{\bP}_{V_0} f(v) = f(v)$ for $v\in V_0$ and $0$ on $V\backslash V_0$. A \emph{semi shift invariant filter} of degree $d$ is a composition $\overline{\bP}_{V_0}\circ Q_d(\bS)$ for a polynomial $Q_d$ of degree $d$. 
\end{Definition}

From \cref{defn:lvs}, the output of a semi shift invariant filter is supported on $V_0$. More generally, we have the following.

\begin{Definition}\label{defn:glvs}
For $k\geq 1$, let $C = (V_1,\ldots, V_k)$ be a tuple of subsets of vertices in $V$ and $D = (d_1,\ldots,d_k)$ be a tuple of non-negative integers each smaller than $n$. The space of semi shift invariant filters $\scJ_{C,D}$ on $C$ of degree type $D$ is the span of semi shift invariant filters supported on $V_j$ of degree $d_j$ for $1\leq j\leq k$, i.e., if $\bF\in\scJ_{C,D}$, then 
\begin{align*}
\bF = \sum_{j=1}^k \overline{\bP}_{V_j} \circ Q_{d_j}(\bS)
\end{align*}
for some polynomials $Q_{d_1},\ldots,Q_{d_k}$.
\end{Definition}

For any $C$ and $D$, $\scJ_{C,D}$ is a vector space, and we use $\dim \scJ_{C,D}$ to denote its dimension. We now give some examples.

\begin{Example} \label{eg:ica}
\begin{enumerate}[a)]
\item Suppose $\bS$ does not have repeated eigenvalues. If $C = V$ and $D = n-1$, then $\scJ_{C,D}$ is the space of all shift invariant filters. 

\item \label{it:FCDex}
We work out a numerical example with the directed cycle graph shown in \figref{fig:ssp1}. The directed adjacency matrix $\bA_G$ is used as the shift operator for the ambient graph $G$. Let $V_0=\{v_0,v_2,v_2,v_5\}$, $V_1 = \{v_1,v_2\}$ and $V_2 = \{v_0,v_5\}$. Set $C = (V_1,V_2)$ and $D=(1,3)$. An example of a filter in the filter family $\scJ_{C,D}$ is given by 
\begin{align*}
\bF = \overline{\bP}_{V_1}\circ \bA_G + \overline{\bP}_{V_2}\circ \bA_G^3.
\end{align*}
For any signal $f$, applying $\bF$ to $f$ permutes its $V_0$ components in the cyclic order according to their positions on the induced subgraph $H_0$. To see this, let $u \leadsto v$ denote the permutation of the $u$ component of a signal to the $v$ component after applying an operator. In $\bF$, the operator term $\overline{\bP}_{V_1}\circ \bA_G$ achieves $v_0 \leadsto v_1 \leadsto v_2$. Here, we notice that $v_2\leadsto v_3$ and so on due to $\bA_G$ but are nulled out by $\overline{\bP}_{V_1}$. Similarly, $\overline{\bP}_{V_2}\circ \bA_G^3$ achieves $v_2\leadsto v_5 \leadsto v_0$. Therefore, in total, $\bF$ results in $v_0\leadsto v_1 \leadsto v_2 \leadsto v_5 \leadsto v_0$.
\begin{figure}[!htb]
\centering
\includegraphics[width=0.5\columnwidth]{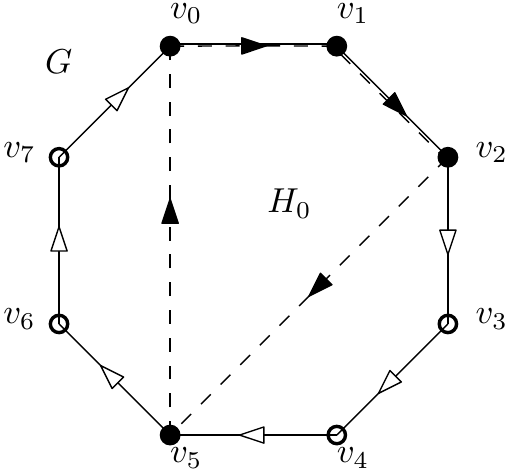}
\caption{An example of semi shift invariant filters on the directed cyclic group with $8$ nodes.}\label{fig:ssp1}
\end{figure}
\end{enumerate}
\end{Example}

Locality of semi shift invariants is explained by the following observation.

\begin{Lemma_A}\label{lem:dhops}
For $V_0 \subset V$ and $d \geq 0$, the $d$-hop neighborhood $B_d(V_0)$ of $V_0$ is the union of vertices that are at most $d$ hops away, in V, from some vertex of $V_0$. Suppose $\bL_G$ is the Laplacian of $G$ and $\bL_{V_0,d}$ is the Laplacian of the induced subgraph on $B_d(V_0)$, extended by $0$ to $V\backslash B_d(V_0)$. Then a semi shift invariant filter $\bF = \overline{\bP}_{V_0} \circ Q_d(\bL_G)$ of degree $d$ is also given by $\bF = \overline{\bP}_{V_0}\circ Q_d(\bL_{V_0,d})$.
\end{Lemma_A}

\begin{IEEEproof}
Let $x$ be any graph signal. As $Q_d$ is of degree $d$, the value of $Q_d(\bL_G)(x)$ at each vertex $v\in V$ depends only on the signal values at vertices at most $d$ hops away from $v$. The filters $Q_d(\bL_G)$ and $Q_d(\bL_{V_0,d})$ are thus the same on $V_0$. As the signal value of $\bF(x)$ is $0$ outside $V_0$, we have the desired identity $\bF = \overline{\bP}_{V_0}\circ Q_d(\bL_{V_0,d})$.  
\end{IEEEproof}

\subsection{Relations among families of semi shift invariant filters}

In filter design or estimation, we usually need to pre-define a filter bank from which the desired filter is chosen. The filter bank should have an appropriate size, measured by for example its dimension. We want the filter bank to be large enough to contain good approximation of the ground truth. On the other hand, a filter bank being too large can hinder effective learning. 

In this paper, we propose to choose $\scJ_{C,D}$ with appropriate $C$ and $D$ as filter bank candidate. The purpose of this subsection is to provide estimation of $\dim \scJ_{C,D}$ and discuss relation among $\scJ_{C,D}$ for different $C$ and $D$.

To motivate our discussion on dimension, we give more examples. For convenience, if $C=\{v\}$ and $D = d$ for some vertex $v$ and non-negative integer $d$, we write $\scJ_{C,D}$ as $\scJ_{v,d}$.

\begin{Example_A} \label{eg:sas}
\begin{enumerate}[a)]
\item \label{it:sld} Suppose a symmetric shift operator $\bS$ does not have repeated eigenvalues, and no eigenvector of $\bS$ has zero components. For each $v\in V$ and $0\leq d\leq n-1$, we have $\dim \scJ_{v,d}=d+1$. To see this, we note that the space $\scJ_{v,d}$ is spanned by $\overline{\bP}_{v}\circ \bS^i$, for $0\leq i\leq d$. Therefore, $\dim\scJ_{v,d} \leq d+1$. We also have $\dim \scJ_{v,d+1} - \dim \scJ_{v,d} \leq 1$. Hence, it suffices to prove the case for $d=n-1$. Suppose $\sum_{0\leq i\leq n-1} a_i \overline{\bP}_{v}\circ \bS^i = 0.$ We write the orthonormal decomposition of  $\bS = \bU\Lambda \bU^{*}$. The diagonal entries of $\Lambda$ are the distinct eigenvalues $\lambda_0,\ldots, \lambda_{n-1}$. Suppose the index of $v$ is $j$. Unwrapping the above identity as a transformation in the frequency domain:
\begin{align*}
\sum_{0\leq i \leq n-1}a_i (\bU_{j,0}\lambda_0^i,\ldots, \bU_{j,n-1}\lambda_{n-1}^{i}) = 0.
\end{align*}
As the row vectors $(\lambda_0^i,\ldots, \lambda_{n-1}^{i})$, $0\leq i\leq n-1$ are linearly independent and the matrix entries $\bU_{j0}, \ldots, \bU_{j,n-1}$ are non-zero, the vectors $(\bU_{j,0}\lambda_0^i,\ldots, \bU_{j,n-1}\lambda_{n-1}^{i})$, $0\leq i\leq n-1$ are linearly independent. Consequently, $a_i=0$ for $0\leq i\leq n-1$.

\item \label{it:sl} Suppose $\bS$ does not have repeated eigenvalues, and no eigenvector of $\bS$ has zero components. If $C = (\{v\})_{v\in V}$ and $D = (n-1,\ldots, n-1)$, then $\scJ_{C,D}$ is the space of all filters on $\mathbb{R}^{n}$. To see this, we note that the dimension of the space of all linear filters is $n^2$. For any vertices $u, v\in V$ and polynomials $Q$ and $Q'$, $\overline{\bP}_{u}\circ Q(\bS)$ and $\overline{\bP}_{v}\circ Q'(\bS)$ are supported on different vertices and are thus linearly independent in $\scJ_{C,D}$. It suffices to show that for each $v\in V$, $\dim \scJ_{v,n-1} = n$, but this follows from \ref{it:sld} in this example. More generally, if $D = (l,\ldots,l)$ for some $l\leq n-1$, then $\scJ_{C,D}$ is the space of \emph{node-variant graph filters} up to degree $l$ described in \cite{seg17}. 
\end{enumerate}
\end{Example_A}

While the trivial space is an extreme case, \cref{eg:sas}\ref{it:sl} is the other extreme. For our applications in the paper, we are also interested in other intermediate cases. The following discussion can be technical, and readers may skip until \figref{fig:lattice} upon first reading. The insight is that if $ C'$ is ``finer'' than $C$ and $D$ is ``smaller'' than $D'$, then $\scJ_{C,D}$ is a subspace of $\scJ_{C',D'}$. The example shown in \figref{fig:lattice} illustrates the general picture.

To state the general result rigorously, we need to introduce more concepts. 

\begin{Definition_A} \label{def:cic}
The tuple $C = (V_1,\ldots, V_k)$ is called \emph{essential} if for each $1\leq i\leq k$, $V_i \backslash \bigcup_{1\leq j\leq k, j\neq i}V_j \neq \emptyset,$
i.e., none of $V_i$ is completely contained in the union of the remaining sets of $C$.
\end{Definition_A}

\begin{figure}[!htb]
\centering
\includegraphics[width=0.9\columnwidth]{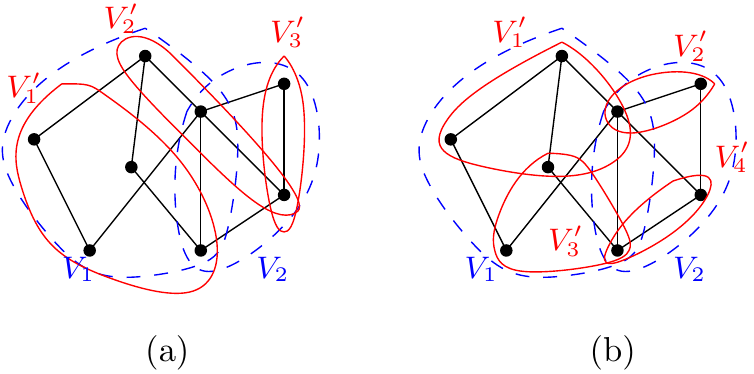}
\caption{In (a), $C= (V_1,V_2)$ (blue) and $C' = (V_1', V_2', V_3')$ (red). $C'$ is not a refinement of $C$ because $V_2'$ is contained in neither $V_1$ nor $V_2$. In (b), $C= (V_1,V_2)$ (blue) and $C' = (V_1', V_2', V_3', V_4')$ (red). Though \cref{def:cic}\ref{it:evi} is satisfied, $C'$ is not a refinement of $C$ because $V_1', V_3'$ are both in $V_1$ but $V_1'\cap V_3' \neq \emptyset$.}
\label{fig:ssp5}
\end{figure}

\begin{Definition_A}\label{def:refinement}
A tuple $C' = (V_1',\ldots, V_{l}')$ is said to be a \emph{refinement} of $C$ if the following holds (cf.\ \cref{fig:ssp5}):
\begin{enumerate}[a)]
\item $\bigcup_{1\leq i\leq k}V_i = \bigcup_{1\leq j\leq l}V_j'$.
\item \label{it:evi} Each $V_j'$ is contained in some $V_i$.
\item If distinct $V_{j_1}', V_{j_2}'$ are contained in the same $V_i$ for some $i$, then $V_{j_1}'\cap V_{j_2}' = \emptyset$.
\end{enumerate}
\end{Definition_A}

We now state our main structural result on $\scJ_{C,D}$ for various tuples $C$ and $D$, whose proof is technical and can be found in \cref{sec:pro}. We compute the dimension of $\scJ_{C,D}$ based on conditions on $C$ and $D$. 

\begin{Proposition_A} \label{thm:ivv}
If $V_1 \subset V_2 \subset V$ and $0\leq d_1 \leq d_2 \leq n-1$, then $\dim \scJ_{V_1,d_1} \leq \dim \scJ_{V_2,d_2}$. Furthermore, suppose $\bS$ does not have repeated eigenvalues, and no eigenvector of $\bS$ has zero components. Then, the following holds.
\begin{enumerate}[a)]
\item\label{it:ivva}  If $C = (V_1,\ldots, V_k)$ is essential and $D = (d_1,\ldots,d_k)$, then $\dim \scJ_{C,D} = \sum_{1\leq i\leq k} d_i +k$.

\item Let $C = (V_1,\ldots, V_k)$, $D=(d_1,\ldots,d_k)$, $C' = (V_1',\ldots, V_l')$ and $D' = (d_1',\ldots, d_l')$. Suppose $C'$ is a refinement of $C$. If $D$ and $D'$ satisfy $d_i\leq d_j'$ whenever $V_j'\subset V_i$, then $\scJ_{C,D} \subset \scJ_{C',D'}$. On the other hand, suppose that $\scJ_{C,D} \subset \scJ_{C',D'}$ and $\bigcup_{1\leq j \leq l} V_j'\subset \bigcup_{1\leq j \leq k} V_j$. If $D$ and $D'$ are constant tuples with $d_i=d_l'=d$ for all $i, l$ and the same $0\leq d <n$, and $C'$ is essential, then $C'$ is a refinement of $C$. 
\end{enumerate}
\end{Proposition_A}

\subsubsection*{Discussions} 
\begin{figure}[!htb]
\centering
\includegraphics[width=0.95\columnwidth]{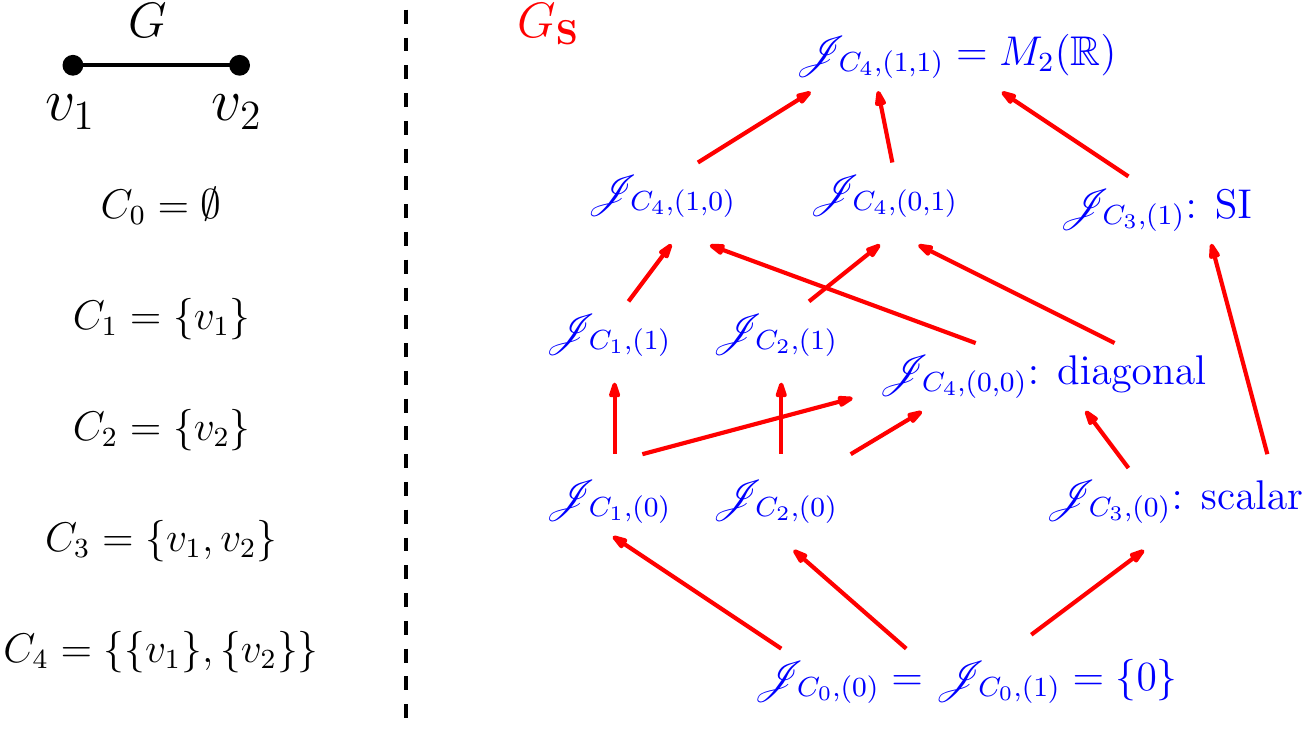}
\caption{In the illustration, the graph $G$ with $2$ nodes $v_1,v_2$ are considered. For $\bS = \bL_G$, we list down all $\scJ_{C,D}$. Their relations are shown in the graph $G_{\bS}$ with each filter bank containing the one below that connected to it by an edge. We also indicate some of the well known filter banks. For example, the join of shift invariant and $\scJ_{C_1,(1)}$ is $M_2(\mathbb{R})$ and their meet is trivial. Hence, these two filter banks are very different from each other. If the true filter is from $\scJ_{C_1,(1)}$, then a learning method using the shift invariant filter bank can be erroneous.}
\label{fig:lattice}
\end{figure}

With the above results, we gain more understanding of different families of semi shift invariant filters. For a chosen $\bS$, it is possible to visualize the relations between different $\scJ_{C,D}$ using a directed acyclic graph $G_{\bS}$, by connecting two such filter banks if one is a maximal subspace of the other (cf.\ \figref{fig:lattice}). More specifically, we associate with each node of $G_{\bS}$ a filter space $\scJ_{C,D}$. Two distinct filter spaces $\scJ_{C_1,D_1}$ and $\scJ_{C_2,D_2}$ are connected by a directed edge in $G_{\bS}$ if and only if the following holds:
\begin{enumerate}[a)]
    \item $\scJ_{C_1,D_1} \subset \scJ_{C_2,D_2}$, and
    \item if $\scJ_{C_1,D_1} \subset \scJ_{C_3,D_3} \subset  \scJ_{C_2,D_2}$, then either $\scJ_{C_1,D_1} = \scJ_{C_3,D_3}$ or $\scJ_{C_3,D_3} =  \scJ_{C_2,D_2}$.
\end{enumerate}

In the graph $G_{\bS}$, we have nodes associated with a large array of candidate filter banks that are between the extreme cases $M_n(\mathbb{R})$ and the trivial space. We may also use concepts in lattice theory to enhance our understanding. For example, $\scJ_{C_1,D_1}$ is an upper bound of $\scJ_{C_2,D_2}$ if $\scJ_{C_1,D_1} \supset \scJ_{C_2,D_2}$ and $\scJ_{C_2,D_2}$ is called a lower bound of $\scJ_{C_1,D_1}$. The condition is equivalent to having a directed path in $G_{\bS}$ from $\scJ_{C_2,D_2}$ to $\scJ_{C_1,D_1}$. The join of two nodes in $G_{\bS}$ is their least upper bound, and the meet is their largest lower bound. The difference of two filter banks can be inspected on $G_{\bS}$ by how far they are away from their meet and join.

For a signal processing task, in additional to finding the correct graph shift operator $\bS$, we also propose to choose an appropriate node in the graph $G_{\bS}$ such that the associated semi shift invariant filter bank is both economic and expressive enough for the problem. 

In the following sections, we discuss applications with numerical experiments. Emphasis is put on how and why $C$ and $D$ are chosen. 

\section{Subgraph signal processing} \label{sec:subgsp}
One primary goal of introducing semi shift invariant filters is to develop the framework of \emph{subgraph signal processing}, which we describe one aspect in this section. 

Suppose $V_0$ is a subset of $V$ of size $n_0$ and $f \in \mathbb{R}^{n}$ is a signal on $V$. We want to perform signal processing on the \emph{subgraph signal} $f_{V_0} = \bP_{V_0}f$, observed only at $V_0$. We call it \emph{subgraph signal processing}. There can be scenarios where such a framework might be needed. It is possible that full observation of $f$ is impossible, or even the full graph structure is unavailable. For example, in a sensor network, it is possible the overall sensor network is unknown or readings from some sensors are missing, due to reasons such as processing delay \cite{Kha16}, damage, energy conservation \cite{TayTsiWin:J07a}, or lack of access because of privacy \cite{SunTay:J20a,SunTay:J20b,WanSonTay:J21}. As another example, full information of a mega size social network may not be readily available, and one can only perform inference and learning with partial signal as well as graph information. A complete discussion about subgraph signal processing can be found in \cite{Jif21}. Here, we focus on describing filter estimation.

The idea is as follows. We want to estimate a filter $\bF_0 \in M_{n_0}(\mathbb{R})$ to process partially observed subgraph signals $f_{V_0}$. For such a filter $\bF_0$, we impose a few conditions: 
\begin{condition} \label{cond:tac}
\begin{enumerate}[a)]
\item The \emph{algebraic condition}: $\bF_0$ is a symmetric transformation, i.e, $\bF_0 \in S_{n_0}(\mathbb{R})$ where $S_{n_0}(\mathbb{R})$ is the space of $n_0\times n_0$ symmetric matrices.
\item \label{it:geo} The \emph{geometric condition}: $\bF_0$ is close (in an appropriate sense) to a semi shift invariant filter $\bF \in \scJ_{C_{V_0},D_{V_0}}$ for suitably chosen $C_{V_0}$ and $D_{V_0}$.
\item The \emph{data fitting condition}: $\bF_0$ models well the given dataset.
\end{enumerate}
\end{condition}

\begin{figure}[!htb]
\centering
\includegraphics[width=0.8\linewidth]{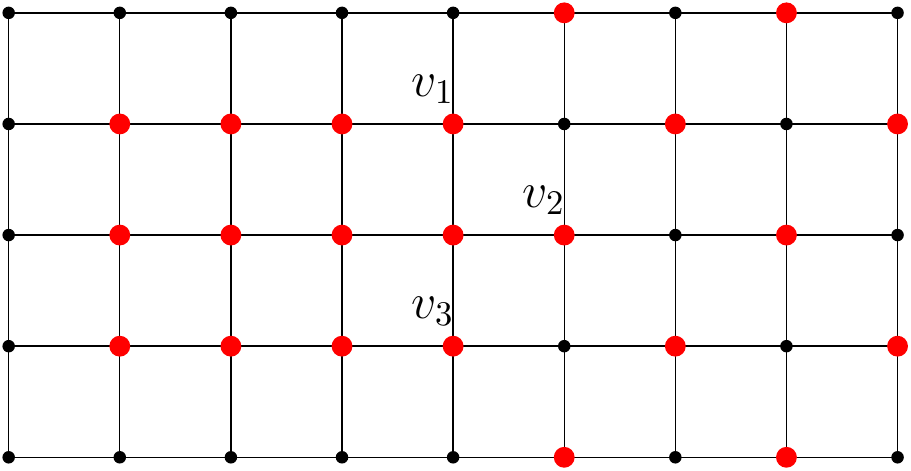}
\caption{Suppose $G$ is the $5\times 9$ lattice graph and $V_0$ consists of the 22 red vertices. The tuple $C_{V_0}= (V_1, V_2)$, where $V_1$ consists of the $13$ vertices, including $v_1,v_2,v_3$ and all the red vertices on their left. $V_2$ has $12$ vertices, including $v_1,v_2,v_3$ and all the red vertices on their right. If $r = 1$, then $D_{V_0} = (2,3)$. All other $V_i$, $i\geq 3$ are empty.}
\label{fig:ssp7}
\end{figure}

Most relevant to the paper is Condition~\ref{cond:tac}\ref{it:geo} the geometric condition. The intuition is that the filter $\bF_0$ we seek should be an approximation of the restriction of a filter $\bF \in M_n(\mathbb{R})$. Recall the primary example for $\bS$ is $\bL_G$. Consider two distinct vertices $v_1, v_2 \in V_0$ being $d$ hops away from each other in $G$. In order to exchange signal information between $v_1$ and $v_2$ in $G$, the $d$-th power of the Laplacian $\bL_G$ is needed. Therefore, locally at each $v \in V_0$, the filter $\bF$ should take the form of a polynomial in $\bL_G$, whose degree depends on how far it is away from other vertices in $V_0$. In view of this, we define $C_{V_0}$ and $D_{V_0}$ as follows. An illustration is given in \cref{fig:ssp7}.

\begin{Definition} \label{defn:gas}
Let $\delta_G$ be the diameter of $G$. Given a subset of vertices $V_0 \subset V$, let $\set[|]{V_i \given 1\leq i\leq \delta_G}$ be a collection of subsets of $V_0$ such that the following holds:
\begin{enumerate}[a)]
\item \label{it:bv} $\bigcup_{1\leq i\leq \delta_G}V_i = V_0$. 
\item \label{it:fei} The set $V_1$ consists of those vertices in $V_0$ that have 1-hop neighbors in $V_0$ as well as these neighbors. For each $i=2,\ldots,\delta_G$, if $v\in V_0$ has $i$-hop (calculated in $G$) neighbors in $V_0$ but no $j$-hop neighbors for $j\leq i-1$ in $V_0$, then $v$ and all of its $i$-hop neighbors are in $V_i$.  
\end{enumerate}
Let $\calI_{V_0} = \set*[|]{i=1,\ldots,\delta_G \given V_i \ne \emptyset}$. As a tuple of subsets of $V_0$, $C_{V_0}$ is formed by including each (non-empty) $V_i$, $i\in\calI_{V_0}$. Correspondingly, fix a small integer $r\geq 0$. As a tuple of degrees, $D_{V_0}$ includes $i+r$ for each $V_i$ in $C_{V_0}$.
\end{Definition}

To elaborate further, for \ref{it:bv} in \cref{defn:gas}, we want the family of filters to perform local operations on $V_0$, therefore we require the sets $\{V_i \mid i\in \calI_{V_0}\}$ cover $V_0$. For \ref{it:fei}, each vertex in $V_i$ finds at least an $i$-hop neighbor. Therefore, to transfer signals in $V_i$, we need to include the $i$-th power of the graph shift operator as discussed earlier. To allow flexibility, we may relax the degree to $i+r$ on each $V_i$ with a small integer hyper-parameter $r\geq 0$ (hence of order $\calO(1)$\footnote{$\calO(\cdot)$ stands for the big-O notation.} as compared with the size of the graph) to allow tuning and adjustment. In summary, a $\bF\in\scJ_{C_{V_0},D_{V_0}}$ has the form
\begin{align}\label{eq:F}
	\bF = \sum_{i\in\calI_{V_0}} \overline{\bP}_{V_i}\circ Q_{i+r}(\bS) = \sum_{i\in\calI_{V_0}}\sum_{j=0}^{i+r} a_{ij} \overline{\bP}_{V_i} \circ\bS^j.
\end{align}

We now describe explicitly the learning framework by filling in other details of Conditions~\ref{cond:tac}. Suppose we have graph signals $y_t$ and $z_t$ related by $z_t = \widetilde{\bF}(y_t)$, $t=1,\ldots,T$, for some unknown filter $\widetilde{\bF}$. We have access to only the partial observations $x_t=\bP_{V_0}(y_t)$ and $x'_t=\bP_{V_0}(z_t)$, for $t=1,\ldots,T$. Our objective is to estimate a filter $\bF_0$ on $V_0$ to approximate $\widetilde{\bF}$. We propose to solve the following joint optimization problem:
\begin{subequations}\label{prob0}
\begin{align}
\min_{\bF\in \scJ_{C_0,D_0}, \bF_0\in S_{n_0}(\mathbb{R})}\ & \sum_{t=1}^T \norm{x'_t - \bF_0(x_t)}_2^2, \label{prob0_obj}\\
\st\ & \ell(\bP_{V_0}\circ \bF, \bF_0\circ \bP_{V_0})\leq \epsilon, \label{prob0_cons}
\end{align}
\end{subequations}
where $\epsilon$ is a positive constant and $\ell$ is a fixed loss function. 

The filter $\bF_0$ is our target filter and $\bF$ is considered as capturing the effective of $\widetilde{\bF}$ on signals on $V_0$. Moreover, $\ell$ is a loss that measures the difference between $\bP_{V_0}\circ \bF$ and $\bF_0\circ \bP_{V_0}$, and we may choose $\ell$ to be the operator norm or the $L^2$-norm. The loss $\ell$ acts as a regularizer if we consider the Lagrangian form:
\begin{subequations} \label{prob:filter}
\begin{align} \label{eq:mff}
\min_{\bF\in \scJ_{C_0,D_0}, \bF_0\in S_{n_0}(\mathbb{R})}\ & \sum_{t=1}^T \norm{x'_t - \bF_0(x_{t})}_2^2 \\& + \beta \ell(\bP_{V_0}\circ \bF, \bF_0\circ \bP_{V_0})^2,
\end{align}
for some positive constant $\beta$. 
\end{subequations}

We note that to solve \cref{prob:filter}, we do not require observation of the full graph signal. In addition, as long as a basis of $\scJ_{C_0,D_0}$ is given, we do not even need the structure of $G$. This is useful when it is impossible to recover the full graph signals $y_t$ and $z_t$ from $x_t$ and $x'_t$, respectively. On the other hand, viewing $x_t$ and $x'_t$ as graph signals independent of $y_t$ and $z_t$, and solving \cref{prob0_obj} without the constraint \cref{prob0_cons} (as is typically done in GSP), ignores the prior knowledge that $x_t,x'_t$ are generated from $y_t,z_t$.

For the rest of the section, we show simulation results. On a given graph $G=(V,E)$, we generate a shift invariant filter $\widetilde{\bF}$ in the form 
\begin{align*}
\widetilde{\bF} = a_0+a_1\bL_G+a_2\bL_G^2, 
\end{align*}
where $\bL_G$ is the graph Laplacian of $G$ and $a_0, a_1, a_2$ are random coefficients chosen uniformly in the interval $[0,1]$. To generate a graph signal $y_t$, for $1\leq t \leq T$, we generate its GFT coefficients uniformly and randomly in the interval $[0,1]$. Let $z_t= \widetilde{\bF}(y_t)$. We assume that we observe only $y_t$ and $z_t$ at a subset of vertices $V_0$. We solve \cref{prob:filter} to obtain the solution $(\bF^*,\bF_0^*)$ by setting $\ell$ as the $L^2$-norm. 

\begin{figure}[!htb]
%\centering
\begin{subfigure}[b]{0.48\textwidth}
%\centering
\includegraphics[width=1\columnwidth, trim=0cm 6cm 0cm 6cm,clip]{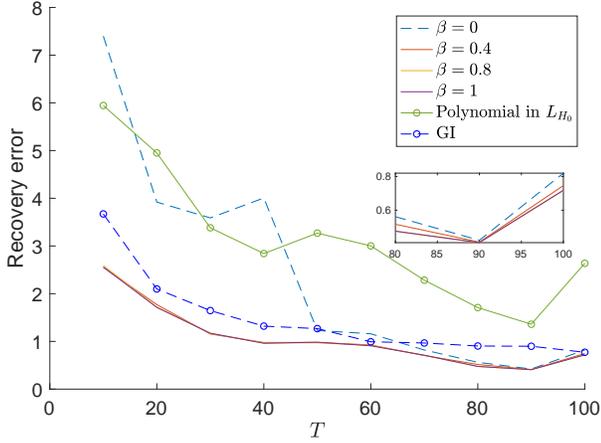}
\caption{The square lattice}
\end{subfigure}
\begin{subfigure}[b]{0.48\textwidth}
%\centering
\includegraphics[width=1\columnwidth, trim=0cm 6cm 0cm 6cm,clip]{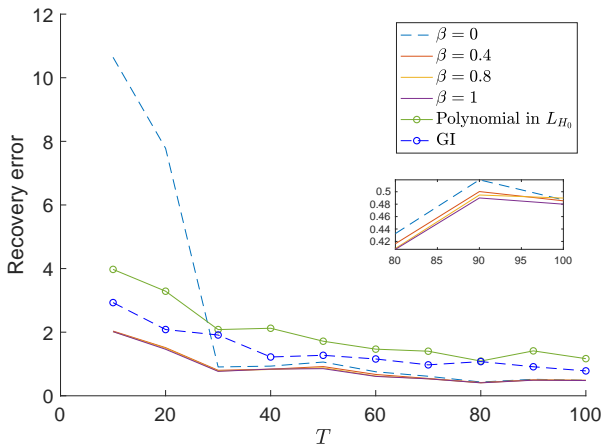}
\caption{The power plant network}
\end{subfigure}
\caption{Performance of filtering learning for the square lattice and the power plant network. We highlighted $\beta=0$ and GI in dashed blue curves and polynomial filter in $\bL_{H_0}$ in dashed red curve. }\label{fig:ssp15}
\end{figure} 

We run $100$ experiments for each set of parameters on the square lattice and the Arizona power network \cite{Ari12} (of size $47$) respectively. The performance is evaluated by computing the average recovery error $\norm{\bP_{V_0}(z_t)-{\bF_0^*}(\bP_{V_0}(y_t))}_2$. For the square lattice, we set $|V_0|\approx 0.4n$; and for the power plant network, we set $|V_0|\approx 0.6n$. We test various $\beta = 0, 0.2, 0.4, 0.6, 0.8, 1$ in \cref{prob:filter} to investigate the contribution of the regularizer given by the loss $\ell$. The case $\beta=0$ corresponds to the case where there is no regularization. We also vary $T$ from $10$ to $100$. 

We consider two baselines for comparison. 
\begin{enumerate}[a)]
    \item The subset of nodes $V_0$ induces a subgraph $H_0 = (V_0,E_0)$ of $G$, where two nodes in $V_0$ is connected by an edge if they are connected by an edge in $G$. We learn $\bF_0^*$ in the form of a polynomial in $\bL_{H_0}$, i.e., $\bF_0^*$ is SI w.r.t.\ $\bL_{H_0}$. 
    \item We compare with a signal interpolation approach called \emph{GI}, that requires full knowledge of the ambient graph $G$. Briefly, given a subgraph signal $x$, we perform an interpolation \cite{Nar13} to obtain a full graph signal $y$. Then we perform signal processing tasks with $y$ and $G$. The filter candidates are polynomials in $\bL_G$ on the reconstructed signals.
\end{enumerate}
 
The results are shown in \cref{fig:ssp15}. We see that by solving \cref{prob:filter} with sufficient regularization, i.e., $\beta\geq 0.4$, the average recovery error is generally smaller. It is also consistently better than GI by an observable margin across the entire range of $T$. The effect of regularization is more prominent when $T$ is small. Furthermore, different $\beta\geq 0.4$ have almost identical performance. Still, if we zoom in ($T=80,90,100$), we see that larger $\beta$ yields a slightly better performance. We observe that our approach gives a much better performance than attempting to learn the filter using $\bL_{H_0}$.

\section{Graph neural networks} \label{sec:gnn}
In this section, we discuss applications of semi shift invariant filters in graph neural networks (GNNs). We first give a brief overview of GNNs.

GNNs are neural network models that operate on graph structured data. By inputting a graph $G$ and a matrix of node features $\bX$ to a GNN model, embeddings of graph components such as nodes and edges can be learned for downstream tasks through neighborhood aggregation process \cite{homophilyma2022is}. In particular, we want to study semi-supervised node classification problems. Each node of the graph has a class label such as article type in the citation network Citeseer. We want to estimate the class label of every node, assuming only the class labels of a small percentage of nodes are observed. We consider both homogeneous and heterogeneous graphs. A homogeneous graph is a graph with a single node type and a single edge type, while a heterogeneous graph has multiple node and/or edge types. 
One of the most conventional GNN models is graph convolutional network (GCN) \cite{kipf2017semi}. In the nutshell, GCN uses convolutional aggregations and is made up of multiple graph convolutional layers. Each graph convolutional layer is of the form:
\begin{equation}
\bH^{(l+1)} = \sigma(\tilde{\bD}^{-\frac{1}{2}}\overline{\bA}_G\tilde{\bD}^{-\frac{1}{2}}\bH^{(l)}\bW^{(l)}). 
\end{equation}
In the expression, $\sigma$ is an activation function, and $\overline{\bA}_G = \bA_G + \bI_n$ is the adjacency matrix with self-loops. The degree matrix $\tilde{\bD}$ of $\overline{\bA}_G$ contains information about the number of edges attached to each node, and $\bW^{(l)}$ is the trainable layer-specific weight matrix. The initial $\bH^{(0)}$ is taken to be $\bX$. 

Most of subsequent GNN models inherit the same philosophy of GCN that performing. Examples include the graph attention model GAT \cite{velickovic2018graph} that involves an additional the step of estimating edge weights and the hyperbolic model HGCN \cite{hgcn2019} that performs a hyperbolic version of convolutional aggregation. Moreover, heterogeneous GNN model HAN \cite{han2019} that generates homogeneous graphs from  heterogeneous graphs using the concept of meta-path. In the subsequent subsections, we propose to take a slightly different approach that performs aggregation that may change for different (cluster of) nodes, in the spirit of the paper.

\subsection{Homogeneous GNN}

\subsubsection{Node homophily}
To motivate, we consider the concept of node homophily. Even though GNN models have achieved prominent results for graph learning tasks, the neighborhood aggregation mechanism has implicit graph homophily assumption, where ``similar" nodes are connected to one another \cite{li2022findinghomophily}. When the graph is heterophilous i.e. many neighboring nodes belonging to dissimilar classes, the models tend to perform poorly \cite{twosideshomophily}. Many works then design various graph filters to address heterophily \cite{twosideshomophily,fagcn2021}. Nevertheless, they neglect the fact that a graph can have different extend of homophily/heterophily at different regions of the graph and a learned filter is still indifferently applied over the entire graph. In this subsection, we investigate the homophily of classes of nodes. In the next subsection, based on the observations, we modify GCN which learns a first order shift invariant filter expressed as the adjacency matrix $\bA_G$ to instead learn semi-shift invariant filters that are restricted to different subsets of vertices in the graph. The new model shall be called \emph{SemiGCN}. 

We want to investigate the extend of node homophily in different classes. Node homophily is typically defined based on similarity of connected node pairs. Nodes are deemed to be similar if they are of the same node label. We slightly modify the homophily metric introduced by \cite{Pei2020GeomGCN} to be $\eta$ class specific as follows:   
\begin{equation}
\resizebox{\linewidth}{!}{
    $\displaystyle
    H_{\eta}(G) = \frac{1}{|V_{\eta}|}\sum_{v \in V_{\eta}}\frac{\text{number of } v\text{'s neighbors of the same label}}{\text{number of } v\text{'s neighbors}}.
    $}
\end{equation}

\begin{figure}[!htb]
\centering
\includegraphics[width=1\columnwidth]{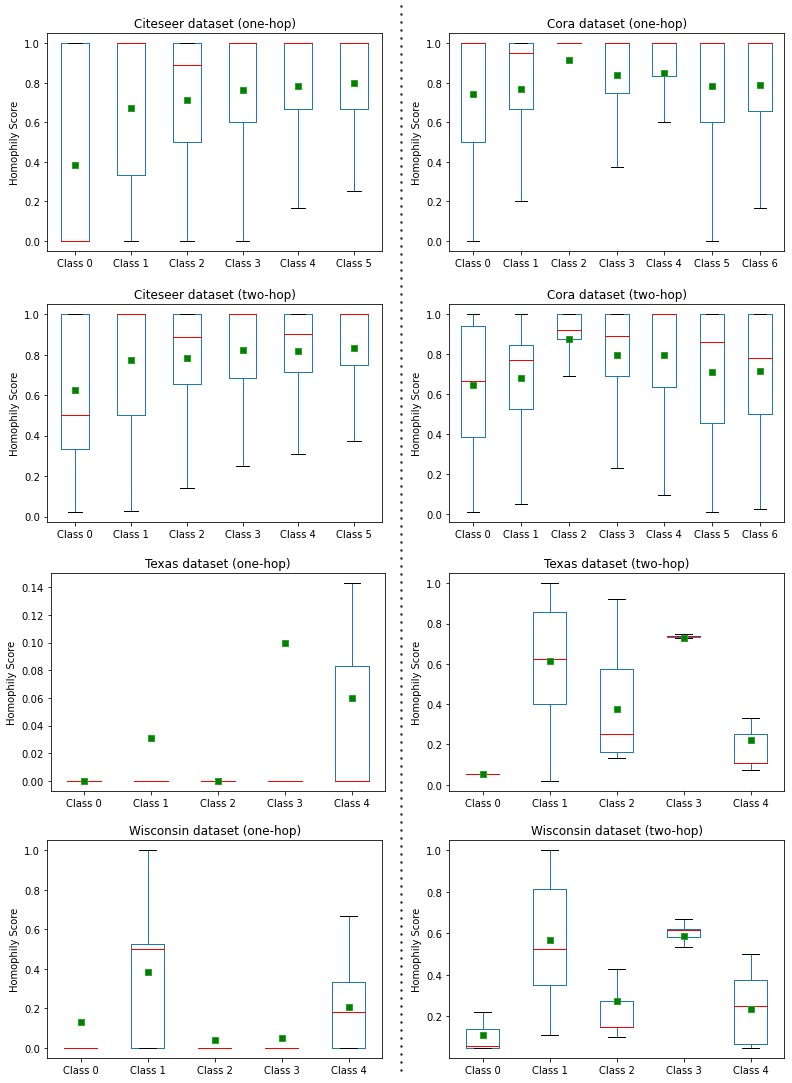}
\caption{One-hop and two-hop homophily score of homogeneous datasets. Best viewed in color; Green squares represents average score and red line denotes median score.}\label{fig:homophily_score}
\end{figure}

A large $H_{\eta}(G)$ implies that the homophily of that specific class is strong. We plot the class specific homophily scores of four homogeneous graph datasets as seen in \cref{fig:homophily_score}. We observed that even though Citeseer and Cora datasets are generally deemed highly homophilic while Texas and Wisconsin datasets heterophilic\cite{Pei2020GeomGCN,homophilyma2022is}, the statistics of homophily score across the classes for the datasets are not necessarily uniform. Thus, it is not ideal to learn and utilise only a single filter across the entire graph.

\subsubsection{SemiGCN}
In this subsection, we introduce SemiGCN which employs semi-shift invariant filters. Recall that a semi-shift invariant filter belongs to a filter bank $\scJ_{C,D}$, for appropriately chosen $C = (V_1,\ldots, V_k)$ and $D = (d_1,\ldots, d_k)$ (cf.\ \cref{defn:glvs}). A layer of semi-shift graph convolutional layer for $k\geq1$ subsets, where $k$ is a hyperparameter to be set, has the following layer-wise propagation rule:

\begin{equation} \label{eq:semi}
    \bH^{(l+1)} = \sigma(\sum_{j=1}^{k}\sum_{i=0}^{d_j} \overline{\bP}_{V_j}\circ\widetilde{\bA_G}^{i}\bH^{(l)}\bW^{(l)}_{i}),
\end{equation}
where $\widetilde{\bA_G}=\tilde{\bD}^{-\frac{1}{2}}\overline{\bA}_G\tilde{\bD}^{-\frac{1}{2}}$ is the normalised adjacency matrix and $d_j$ is the subset-specific polynomial degree. In practice, constraining the number of parameters learned can be beneficial as it addresses overfitting issues. Hence, the trainable degree-specific weight matrix $\bW_{i}$ is shared among subsets if the subsets are tuned to include that order. With the new layer-wise propagation rule, we then test our model in semi-supervised node classification task on homogeneous graphs, as well as heterogeneous graphs in \cref{sec:het}.
% GCN serves as our baseline model for comparison since the formulation of GCN learns a first order shift invariant filter expressed as the adjacency matrix $\bA_G$ that is shared over the whole graph.

Unless otherwise mentioned, we train a two-layer GCN as our baseline, a two-layer SemiGCN for our model and evaluate prediction accuracy on the test set. The hidden units are set to 16 for homogeneous graph datasets and 64 for heterogeneous graph datasets in \cref{sec:het}. All models were implemented using the DGL\cite{wang2019dgl} library. Other hyperparameters are tuned to yield best performance. 

We evaluated our model on four homogeneous datasets. Two of them are citation networks: Cora and Citeseer while the other two are web page datasets: Texas and Wisconsin. The citation networks have nodes representing documents and edges denoting citation links while the web page datasets have nodes as web pages and edges as hyperlinks. The statistics of datasets are as in \cref{table:homo_dataset_stats}.

\begin{table}[!htb]
\captionsetup{justification=centering}
\caption{Summary of homogeneous graph datasets}
\label{table:homo_dataset_stats}
\centering
\resizebox{\columnwidth}{!}{
\begin{tabular}{ccccc} 
\toprule
\multicolumn{1}{c}{Dataset} & \multicolumn{1}{c}{\# Nodes} & \multicolumn{1}{c}{\# Edges} &{\# Classes}& {\# Features} \\ 
\midrule
Cora                        & 2708                        & 5429                          &  7            & 1433\\
Citeseer                    & 3327                        & 4732                          &  6            & 3703\\
Texas                       & 183                         & 309                           &  5            & 1703\\
Wisconsin                   & 251                         & 499                           &  5            & 1703\\
\bottomrule
\end{tabular}}
\end{table}

As one of the most important steps, we describe how the tuple of vertex sets $C= (V_1,\ldots, V_k)$ is chosen. Essentially, we apply an unsupervised clustering algorithm, K-means\cite{kmeans1056489}, on the node features $\bX$ with the number of clusters $k$ in $C$ set to the number of classes in the dataset. Let $\bX_i$ be the $i$-th row of $\bX$, which is the feature vector of $i$-th node $v_i$. The algorithm randomly initialises $k$ cluster centroids and then iteratively updates the centroids such that the intra-cluster squared euclidean distance is minimized \cite{kmeans2010}. The objective function is given by $\sum_{i=1}^{k}\sum_{v_j \in V_i}\lVert \bX_j - \bX_{c_i} \rVert^2$ where $v_{c_i}$ refers to the centroid of $V_i$. A node is deemed to be in the cluster of its nearest cluster centroid. We remark that instead of using K-means, it is possible to use other clustering approach such as applying a pre-trained GCN model. Once $C$ is obtained, we apply (\ref{eq:semi}) for the SemiGCN layers with $D$ tuned as hyperparameters. 

The experimental results are as shown in \cref{table:cls_result_homo}. We see that for the heterophily datasets (Wisconsin and Texas), our approach has significant improvements over popular models such as GCN and GAT. Even for homophily datasets such as Citeseer and Cora, the method still demonstrates comparable performance.

%We note that $k$ can be set to other values to simplify hyperparameter tuning later on and other initialisation methods for $C$ are also possible. Potentially, other initialisations might improve our model's performance to a larger extend since K-means only uses node features to determine the subsets but we can ideally leverage on the connectivity information present to obtain better clusters. 

% We initialise the subsets of vertices $C$ in two methods:
% \begin{enumerate}[a)]
%     \item K-means with the number of clusters $k$ set to the number of classes in the dataset. We note that $k$ can be set to other values to simplify hyperparameter tuning later on.
%     \item Classification result from a trained two-layer GCN. The number of subset $k$ in SemiGCN is then naturally the number of classes in the dataset.
% \end{enumerate}

% The method for setting the subsets are denoted in brackets after the model name. From the empirical results depicted in \cref{table:cls_result_homo}, we observe that initialising $C$ using K-means gives sub-optimal result as compared to the second method. Possibly since K-means uses only node features to learn the clusters whereas a trained GCN's classification also separates the nodes into subsets based upon their connectivity.

\begin{table}[!htb]
\captionsetup{justification=centering}
\caption{Node classification results on homogeneous graph in terms of accuracy (\%). Averaged over ten runs, best performance boldfaced.}
\label{table:cls_result_homo}
\centering
\resizebox{1\linewidth}{!}{
\begin{tabular}{@{}cccc@{}}
\toprule
Datasets        & GAT              & GCN                    & SemiGCN (K-means)   \\ \midrule
Wisconsin       & 56.42 $\pm$ 1.65 & 53.49 $\pm$ 1.96       &  \bf{89.31 $\pm$ 1.09}        \\
Texas           & 55.26 $\pm$ 3.72 & 55.39 $\pm$ 4.56       &  \bf{92.11 $\pm$ 2.64}     \\
Citeseer        & 71.29 $\pm$ 0.55 & 70.81 $\pm$ 0.86       &  \bf{71.77 $\pm$ 0.30}       \\
Cora            & \bf{82.42 $\pm$ 0.98} & 81.02 $\pm$ 0.89  & 80.98 $\pm$ 0.78 \\
\bottomrule
\end{tabular}}
\end{table}
    
\subsection{Heterogeneous GNN} \label{sec:het}
\begin{figure}[!htb]
\centering
\includegraphics[width=0.7\columnwidth]{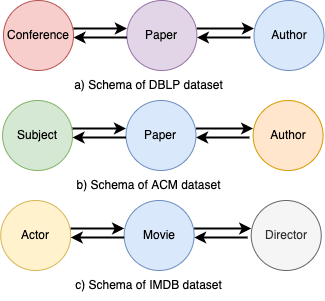}
\caption{Schema of heterogeneous graph datasets. a) DBLP b) ACM and c) IMDB. The node type colored in blue are the nodes to classify in the node classification task. Best viewed in color.}\label{fig:heterograph}
\end{figure}

In this subsection, we consider heterogeneous graphs. These graphs have innate schema as seen in \cref{fig:heterograph}. Hence, they are more complex than homogeneous graphs and require mechanisms catering to the node and/or edge types to achieve state-of-the-art results. 

The node classification task is performed using three heterogeneous benchmark datasets where two of them are citation networks DBLP and ACM and the third is a movie dataset IMDB. Characteristics of the heterogeneous graph datasets are summarised in \cref{table:dataset_stats}. 

The inherent heterogeneity allows us to construct $C = (V_1,\ldots, V_k)$ (used for (\ref{eq:semi})) in a natural way. Since multiple node types and edge types give rise to a non-homogeneous situation, the subsets could be dependent on these characteristics. We group the nodes according to edge types.
More specifically, assuming their are $k$ different types of edges, then $V_i$ contains all the nodes who are end nodes of edges of $i$-th type. As a consequence, a node can appear in multiple $V_i$. Taking IMDB as an example, since actor and director nodes can only be connected to movie nodes, we place movie nodes and actor nodes into one subset and movie nodes and director nodes into another subset. The degree combinations $D$ are tuned to $(2,1)$ for ACM, $(3,1)$ for IMDB and $(3,2)$ for DBLP. A two-layer SemiGCN was applied for DBLP while only one-layer was employed for ACM and IMDB. 

\begin{table}[!htb]
\captionsetup{justification=centering}
\caption{Summary of heterogeneous graph datasets}
\label{table:dataset_stats}
\centering
\resizebox{\columnwidth}{!}{
\begin{tabular}{cccccc} 
\toprule
\multicolumn{1}{c}{Dataset} & \multicolumn{1}{c}{\# Nodes} & \multicolumn{1}{c}{\# Edges} & {\# Node type} &{\# Classes}& {\# Features} \\ 
\midrule
DBLP                        & 18405                        & 67946                        & 3                &  4            & 334\\
ACM                         & 8994                         & 25922                         & 3               &  3            & 1902\\
IMDB                        & 12772                        & 37288                        & 3                &  3            & 1256\\
\bottomrule
\end{tabular}}
\end{table}

We compare with both GCN itself and HAN\cite{han2019}. The results are as shown in \cref{table:cls_result_hetero}. We note that GCN is not designed for heterogeneous graphs. Nevertheless, its modification SemiGCN performs significantly better than GCN in this task. This shows the benefit of employing semi-shift invariant filters in situations where different way of information gathering is required across different subsets. Moreover, SemiGCN even outperforms HAN, which is dedicated to handle heterogeneous graphs. This maybe due partially to the fact that SemiGCN is designed to remove inherent homogeneous assumptions in existing models such as GCN.  

\begin{table}[!htb]
\captionsetup{justification=centering}
\caption{Node classification result on heterogeneous datasets (Standard split). Averaged over ten runs, best performance boldfaced.}
\centering
\resizebox{0.95\columnwidth}{!}{
\begin{tabular}{@{}ccccc@{}}

\toprule
Datasets                    & Metrics  & HAN               & GCN              & SemiGCN        \\ \midrule
\multirow{2}{*}{DBLP}       & Macro-F1 & 91.93 $\pm$ 0.27  & 87.65 $\pm$ 0.29 & \bf{94.10 $\pm$ 0.43}  \\
                            & Micro-F1 & 92.51 $\pm$ 0.24  & 88.71 $\pm$ 2.74 & \bf{94.81 $\pm$ 0.39}  \\ \hline
\multirow{2}{*}{ACM}        & Macro-F1 & 92.01 $\pm$ 0.76  & 91.46 $\pm$ 0.48 & \bf{92.06 $\pm$ 0.33} \\
                            & Micro-F1 & 90.93 $\pm$ 0.73  & 91.33 $\pm$ 0.47 & \bf{91.98 $\pm$ 0.34}  \\ \hline
\multirow{2}{*}{IMDB}       & Macro-F1 & 56.56 $\pm$ 0.77  & 56.72 $\pm$ 0.49 & \bf{58.56 $\pm$ 0.69}  \\
                            & Micro-F1 & 57.83 $\pm$ 0.93  & 58.31 $\pm$ 0.51 & \bf{60.10 $\pm$ 0.70} \\ \bottomrule
\end{tabular}}
\label{table:cls_result_hetero}
\end{table}

\section{Conclusions} \label{sec:con}

In this paper, we generalize the notion of shift invariant filters in GSP by introducing semi shift invariant filters. We study their properties and demonstrate how they can be used in signal processing and machine learning applications. 

\appendices

\section{Proofs of \cref{thm:ivv}}\label[Appendix]{sec:pro}

\begin{IEEEproof}
As $V_1\subset V_2$ and $d_1\leq d_2$, any $\bF\in \scJ_{V_1,d_1}$ is the projection $\overline{\bP}_{V_1}\circ \bF'$ of some $\bF' \in \scJ_{V_2,d_2}$ to $V_1$. Furthermore, if $\bF_1',\ldots, \bF_k' \in \scJ_{V_2,d_2}$ are filters such that $\overline{\bP}_{V_1}\circ \bF_1',\ldots, \overline{\bP}_{V_1}\circ \bF_k'$ are linearly independent in $\scJ_{V_1,d_1}$, then $\bF_1',\ldots, \bF_k'$ are linearly independent in $\scJ_{V_2,d_2}$ since the sum and projection operations commute. Hence, a basis of $\scJ_{V_1,d_1}$ consists of the projection of linearly independent filters in $\scJ_{V_2,d_2}$, whence $\dim \scJ_{V_1,d_1}\leq \dim \scJ_{V_2,d_2}$. We next further assume that $\bS$ does not have repeated eigenvalues, and no eigenvector of $\bS$ has zero components.

For a), as $C$ is essential, each $V_i$ contains a vertex $v_i$ outside every $V_j, j\neq i$. By \cref{eg:ica}\ref{it:sld} and the above result, we have 
\begin{align*}
d_i+1\leq \dim \scJ_{v_i,d_i} \leq \dim \scJ_{V_i,d_i} \leq d_i+1.
\end{align*}
Therefore, we must have $\dim \scJ_{v_i,d_i} = \dim \scJ_{V_i,d_i}$, and the (surjective) projection $\overline{\bP}_{v_i}: \scJ_{V_i,d_i} \to \scJ_{v_i,d_i}$ is an isomorphism. For each $i$, let $C_{-i}$ be obtained from $C$ by removing $V_i$ and $D_{-i}$ be obtained from $D$ by removing $d_i$. If $\bF \in \scJ_{v_i,d_i}$ and $\bF'\in \scJ_{C_{-i},D_{-i}}$ where both $\bF$ and $\bF'$ are not trivially $0$, then they are linearly independent of each other. Indeed, if $a\bF+b\bF'=0$, then applying $\overline{\bP}_{v_i}$, we have $a\overline{\bP}_{v_i}\circ \bF=0 \in \scJ_{v_i,d_i}$. As $\overline{\bP}_{v_i}\circ \bF$ is non-zero, we must have $a=0$ and hence $b=0$. Consequently, filters in $\scJ_{V_i,d_i}$, for $1\leq i \leq k$, are all linearly independent and 
\begin{align*}
\dim \scJ_{C,D} = \sum_{1\leq i\leq k}\scJ_{V_i,d_i} = \sum_{1\leq i\leq k}d_i+k.
\end{align*}

For b), if $C'$ is a refinement of $C$, then each $V_i$ in $C$ is a disjoint union of $V_i = \bigcup_{j=1}^{k_i} V_{i_j}'$ with each $V_{i_j}'$ in $C'$. Let $\bF = \overline{\bP}_{V_i}\circ \sum_{0\leq j \leq d_i}a_j \bS^j \in \scJ_{V_i,d_i}$, where $a_j$, $0\leq j\leq d_i$, are scalars. Then since $d_i \leq d_{i_j}'$, we have 
\begin{align*}
\bF = \sum_{1\leq j\leq k_i} \Big( \overline{\bP}_{V_{i_j}'}\circ \sum_{0\leq j \leq d_i}a_j\bS^j\Big) \in \scJ_{C',D'}.
\end{align*}

To prove the second claim, we verify each of the conditions in \cref{def:cic}. To show the first condition, suppose $\bigcup_{1\leq i\leq k}V_i$ is not contained in $\bigcup_{1\leq j\leq l}V_j'$. Let $v \in \bigcup_{1\leq i\leq k}V_i \backslash \bigcup_{1\leq j\leq l}V_j'$, and $\bF \in \scJ_{C,D}$ be a filter such that $\bP_{v}\circ \bF$ is non-trivial. Such an $\bF$ exists as $\dim \scJ_{v,d} \geq 1$ by \cref{eg:ica}\ref{it:sld}. However, $\bF \notin \scJ_{C',D'}$ as the projection of any filter of $\scJ_{C',D'}$ to $v$ is trivial. This gives rise to a contradiction.

For the second condition in \cref{def:cic}, we first note that since the eigenvectors of $\bS$ have no zero components, if $\bF=\sum_{0\leq j\leq d}a_j\bS^j \ne 0$, then $\overline{\bP}_v \circ \bF \ne 0$ for all $v\in V$. Suppose without loss of generality that $V_1'$ is not contained in any single $V_i$. As we assume that $C'$ is essential, there is a $v$ contained only in $V_1'$ and not any other $V_j'$, $j\ne1$. Let $V_i$ in $C$ contain $v$, $V_1'\backslash V_i \ne \emptyset$, and 
\begin{align*}
\bF = \overline{\bP}_{V_i}\circ \sum_{0\leq j\leq d}a_j\bS^j \in \scJ_{C,D}
\end{align*}
be a non-zero filter. Since $\scJ_{C,D}\subset\scJ_{C',D'}$, $\bF \in \scJ_{C',D'}$, and by considering the projection to $v$, $\bF$ must have a summand $\bF_1 = \overline{\bP}_{V_1'}\circ \sum_{0\leq j\leq d}a_j\bS^j$. However, the projection of $\bF_1$ to $V_1'\backslash V_i$ is non-zero. For each $v' \in V_1'\backslash V_i$, there must be some other $V_j'$ such that: i) $v'\in V_1' \cap V_j'$, and
ii) $\bF$ has a non-zero summand $\bF_j \in \scJ_{V_j',d}$. For such a $V_j'$, there is a $v_j$ contained exclusively (\gls{wrt} $C'$) in $V_j'$. However, $\overline{\bP}_{v_j}\circ \bF_j \neq 0$ and hence $v_j \in V_i$. This implies that 
\begin{align*}
\bF_j = \overline{\bP}_{V_j'} \circ \sum_{0\leq j\leq d}a_j\bS^j
\end{align*}
since $\bF_j$ and $\bF$ have the same projection to $v_j$. In conclusion, for any $v' \in V_1'\backslash V_i$, there is a positive integer $m$ such that 
\begin{align*}
0 = \overline{\bP}_{{v'}}\circ \bF = m\overline{\bP}_{v'}\circ \sum_{0\leq j\leq d}a_j\bS^j \neq 0,
\end{align*}
which is a contradiction.

For the last condition in \cref{def:cic}, consider any $V_i$ and choose any non-zero filter 
\begin{align*}
\bF \in \scJ_{V_i,d} \subset \scJ_{C,D} \subset \scJ_{C',D'}. 
\end{align*}
For any $V_j'\subset V_i$, there is a $v_j$ contained exclusively (\gls{wrt} $C'$) in $V_j'$. Therefore, $\bF$ has a summand $\overline{\bP}_{V_j'}\circ \bF$. Then for any $v\in V_i$, $\overline{\bP}_{v}\circ \bF$ is the same as $m\overline{\bP}_{v}\circ \bF$, where $m$ is the number of $V_j'\subset V_i$ that contains $v$. Hence, $m=1$. Therefore, for any distinct $V_{j_1}', V_{j_2}' \subset V_i$, we have $V_{j_1}'\cap V_{j_2}' =  \emptyset$. The proof that $C'$ is a refinement of $C$ is now complete.
\end{IEEEproof}

\bibliographystyle{IEEEtran}
\bibliography{IEEEabrv,StringDefinitions,allref}

\end{document}